\documentclass[12pt,preprint]{aastex}
\usepackage{epsfig}
\newcommand{\ec}{\objectname[]{$\eta$~Car}}
\newcommand{\xmm}{{\it XMM-Newton}}
\newcommand{\ch}{{\it Chandra}}
\newcommand\lax{\>\vcenter{\hbox{$<$\hskip-.75em\lower1.0ex\hbox{$\sim$}}}\>}
\newcommand\gax{\>\vcenter{\hbox{$>$\hskip-.75em\lower1.0ex\hbox{$\sim$}}}\>}

\received{}
\begin{document}

\title{X-ray spectroscopy of \objectname[]{$\eta$~Carinae} with {\bf XMM-Newton}}

\author{Maurice A.\,Leutenegger, Steven M.\,Kahn}
\affil{Department of Physics and Columbia Astrophysics Laboratory}
\affil{Columbia University, 550 West 120th Street, New York, NY 10027, USA}
\email{maurice@astro.columbia.edu}
\email{skahn@astro.columbia.edu}
\and
\author{Gavin Ramsay} 
\affil{Mullard Space Science Laboratory}
\affil{University College London, Holmbury St. Mary, Dorking, Surrey, RH5 6NT,
UK}
\email{gtbr@mssl.ucl.ac.uk}

\shorttitle{X-ray spectroscopy of \protect\ec.}
\shortauthors{Leutenegger, Kahn \& Ramsay}
\revised{}
\accepted{}

\begin{abstract}

We present \xmm\ observations of the luminous star
\objectname[]{$\eta$~Carinae}, including a high resolution soft X-ray spectrum
of the surrounding nebula obtained with the Reflection Grating
Spectrometer. The EPIC image of the field around \ec\ shows many early-type
stars and diffuse emission from hot, shocked gas. The EPIC spectrum of the
star is similar to that observed in previous X-ray observations, and requires
two temperature components. The RGS spectrum of the surrounding nebula shows
K-shell emission lines from hydrogen- and helium-like nitrogen and neon and
L-shell lines from iron, but little or no emission from oxygen. The observed
emission lines are not consistent with a single temperature, but the range of
temperatures observed is not large, spanning $\sim\,0.15\,-\,0.6$ keV. We
obtain upper limits for oxygen line emission and derive a lower limit of
$\mathrm{N/O} > 9$. This is consistent with previous abundance determinations
for the ejecta of \ec, and with theoretical models for the evolution of
massive, rotating stars. 

\end{abstract}

\section{Introduction}

The massive, luminous star \objectname[]{$\eta$~Carinae} is famous for an
extended outburst beginning in 1843, during which it temporarily became the
second brightest star in the sky. This outburst gave rise to a bipolar optical
nebula, obscuring the star from direct observation. \ec\ is thought to be
very massive ($M \sim 100~\mathrm{M_\odot}$), and to lose mass at a rate of
$\dot{M} \sim 10^{-3} \, \mathrm{M_\odot~yr^{-1}}$. For a general review of
its history and properties, see \cite*{dav97}.

{\it Einstein} observations of \ec\ showed it to be a complex X-ray source
\citep{sew79,sew82,chl84}. \ec\ has two X-ray emission components: hard,
absorbed ($N_{\mathrm{H}} \sim 5 \, \times 10^{22}\, \mathrm{cm^{-2}}$),
spatially unresolved emission coming from the star, and soft, extended
emission coming from the nebula around the star. The {\it Einstein}
observations also showed that there are many other X-ray sources in the field
around \ec, and that there is diffuse X-ray emission with an extent of about a
degree. 

{\it Ginga} observations found evidence for iron K-shell emission from \ec\
consistent with Fe XXV, indicating a thermal origin for the hard X-ray
emission \citep{koy90}. \cite{cor95} used {\it ROSAT} PSPC observations to
show that the hard X-ray emission is variable.

{\it ASCA} observations obtained much higher quality spectra, and found
evidence for a very strong \ion{N}{7} Ly~$\alpha$ feature, which was
thought to result from the supersolar abundance of nitrogen in the ejecta
\citep{tsu97,cor98}. This also was consistent with previous optical and UV
spectroscopic observations of the ejecta around \ec\ \citep{dav82,dav86}.

Recent \ch\ ACIS-I imaging observations have resolved \ec\ spatially at the
subarcsecond scale \citep{sew01}. The soft X-ray nebula shows complex
structure with several knots of X-ray emission. \ch\ HETGS observations have
given us the first high resolution X-ray spectrum of the star, showing that
the hard emission is non-isothermal, with emission lines from H- and He-like
iron, calcium, argon, sulfur, and silicon \citep{cor01}.

In this paper, we report the results of \xmm\ observations of \ec, including
the high resolution soft X-ray spectrum obtained with the Reflection Grating
Spectrometer (RGS) \citep{dh01}. Until now, no X-ray observatory has been able
to obtain high resolution soft X-ray spectra of extended sources. RGS has a
spectral resolution of about 0.1 \AA\ for the $\sim$ 1' nebula of \ec, or
$\frac{\lambda}{\Delta\lambda} \sim 200$ at 20 \AA. This is important in the
case of \ec, because we can study the physical state of the X-ray nebula in
detail, and obtain much more accurate elemental abundance measurements than
with a CCD spectrometer. We also present the EPIC image of the field and the
CCD spectrum of \ec.


\section{Observation and Data Analysis}
\label{obs}

\ec\ was observed with XMM on 2000 July 27-28 for a total of 50 ks, split
into two nearly consecutive observations. The EPIC-MOS1 \citep{tur01} and
EPIC-pn \citep{str01} cameras were operated in full-frame mode, and MOS2 was
operated in small window mode. All of the EPIC cameras used the thick
filter. Due to the optical brightness of \ec, the Optical Monitor was blocked.

The EPIC data were processed with SAS version 5.3.0, and
the RGS data were processed with a development version of the SAS
(xmmsas\_20011104\_0842-no-aka). Standard event filtering procedures were
followed for RGS and EPIC. Times with high particle background levels were
filtered out, leaving 45.9 ks of usable RGS exposure and 39.5 ks of EPIC
exposure.  

\ec\ is somewhat piled up in MOS1, but not in MOS2 or pn. The
$0.2\,-\,10\,\mathrm{kev}$ count rate of \ec\ in MOS2 was
$2.5\,\mathrm{cts\,s^{-1}}$.  Because it has moderately higher spectral
resolution than pn, we use only MOS2 for spectroscopy. The canned MOS2
response matrix appropriate for standard MOS event grade selection was used
(e.g. $\mathrm{PATTERN}\,=\,0\,-\,12$). A background region as free as
possible of sources and diffuse X-ray emission was chosen for the MOS2
spectral analysis.


\subsection{EPIC spectral and imaging analysis} 

\begin{figure*}[!ht]
  \begin{center}
    \psfig{file=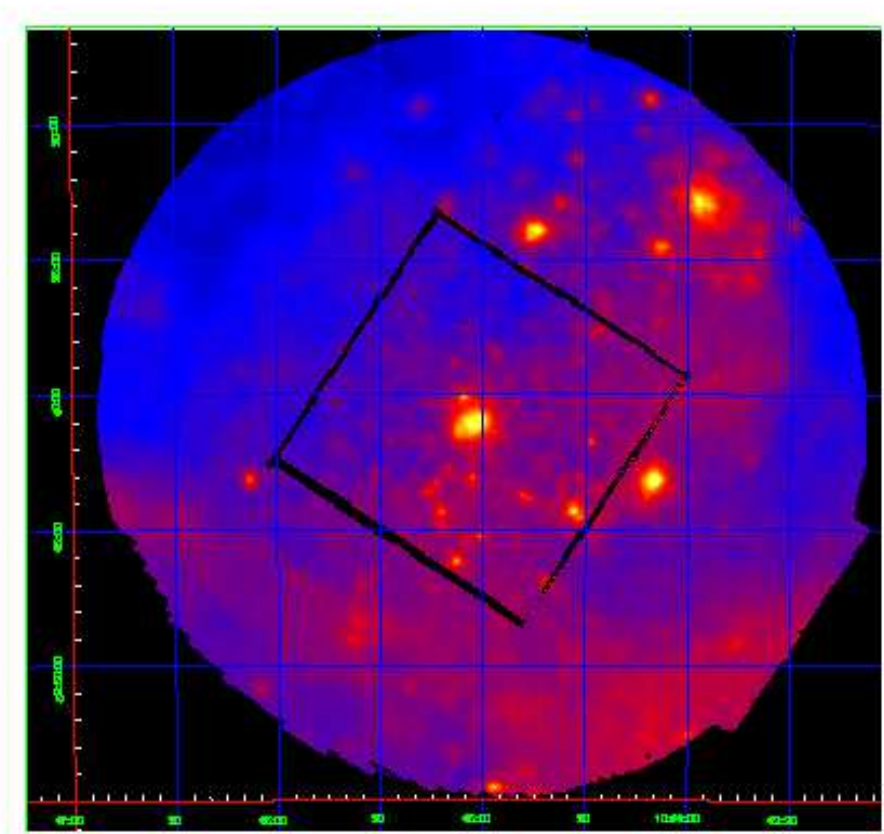, width=4.5in}
  \end{center}
\caption{Combined, exposure corrected EPIC-MOS image of the field around
\protect\ec\ in the energy range 0.3-2.5 keV. The contrast scale is
logarithmic. There is substantial emission from diffuse gas and/or unresolved
point sources over the entire field of view.}
\label{mosimage}
\end{figure*}

The smoothed, exposure corrected EPIC-MOS image in the 0.3-2.5 keV band is
shown in Figure~\ref{mosimage}. It includes data from both MOS cameras.
\ec\ is clearly visible in the center, and several bright stars are also
present, including WR 25 and several O-type stars. There is also substantial
emission from either unresolved point sources or diffuse gas. This emission is
a substantial X-ray (non-particle) background contaminant to the RGS spectrum,
as discussed below.


\begin{figure}[!ht]
  \begin{center}
    \psfig{file=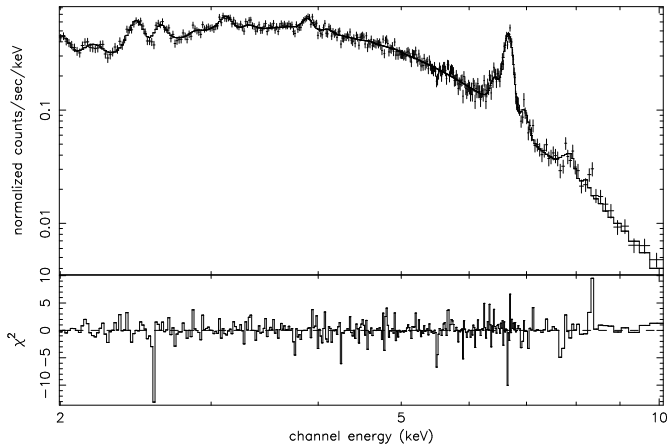, width=2.3in, angle=270}
  \end{center}
\caption{EPIC-MOS2 spectrum of \protect\ec\ with best fit model.}
\label{mosspectrum}
\end{figure}

The EPIC-MOS2 spectrum of \ec\ is shown in Figure~\ref{mosspectrum}. It is
similar to the \ch\ HETGS spectrum \citep{cor01}. There are a number of strong
emission lines, including K-shell lines of hydrogenic and helium-like Si, S,
Ar, Ca and Fe. There is also an iron K fluorescence line from neutral iron in
the optical nebula. As found using the \ch\ HETGS spectrum, a two temperature
model gives a significantly better fit to the EPIC MOS spectrum than a
one temperature model. We only fit the data from $2\,-\,10\,\mathrm{keV}$ in
order to include only emission from the star and to exclude emission from the
nebula. The results of the fit are shown in Table~\ref{tabepic}; abundances
are quoted relative to solar \citep{ag89}. We assume a distance of 2.3 kpc
in determining the luminosity \citep{dav97}.

Table~\ref{tabepic} also shows that Si, S, Ar and Fe are marginally
underabundant, although the relative ratios are close to solar. \cite{cor01}
found that S is marginally overabundant, while Si and Fe were solar within the
errors. However, the inferred abundances may have systematic errors as a
result of fitting a two temperature model to what is likely a continuous
distribution of temperatures (for the both the \ch\ and the EPIC data). 

The unabsorbed luminosity is about the same compared to the epoch of the \ch\
observation. The equivalent width of the neutral iron K fluorescence feature
is 64 eV, compared to 39 eV at the time of the \ch\ observation, but still
lower than the lowest EW observed with ASCA \citep{cor00}.


\begin{table}[h]
\begin{center}
\caption{Best fit parameters for the EPIC-MOS2 spectrum. Abundances are
relative to solar. $L_{X}$ is 2-10 keV. A distance of 2.3 kpc is assumed
\citep{dav97}.}
\begin{tabular}{lr}
\hline
$N_{H,1}$ ($10^{22}\,\mathrm{cm^{-2}}$) & $5.7\,\pm\,0.1$\\
Temperature 1 (keV) & $1.52^{+0.05}_{-0.07}$\\
$L_{X,1}$ (unabsorbed) & $2.5\times10^{34} \, \mathrm{ergs \, s^{-1}}$ \\
$N_{H,2}$ ($10^{22}\,\mathrm{cm^{-2}}$) & $15.2^{+0.73}_{-1.8}$\\
Temperature 2 (keV) & $4.64^{+0.13}_{-0.08}$\\
$L_{X,2}$ (unabsorbed) & $1.2\times10^{35} \, \mathrm{ergs \, s^{-1}}$ \\
Si & $0.6\,\pm\,0.1$\\
S & $0.57\,\pm\,0.06$\\
Ar & $0.61^{+0.12}_{-0.14}$ \\
Ca & $1.1\pm0.2$\\
Fe & $0.59^{+0.02}_{-0.03}$\\
EW Fe (6.4 keV) & $64^{+30}_{-7}$ eV \\
$\chi^{2}_{\nu}$ & 1.11 (351 dof) \\
\hline
\label{tabepic}
\end{tabular}
\end{center}
\end{table}


\subsection{RGS spectral analysis}

\begin{figure*}[!ht]
  \begin{center}
    \psfig{file=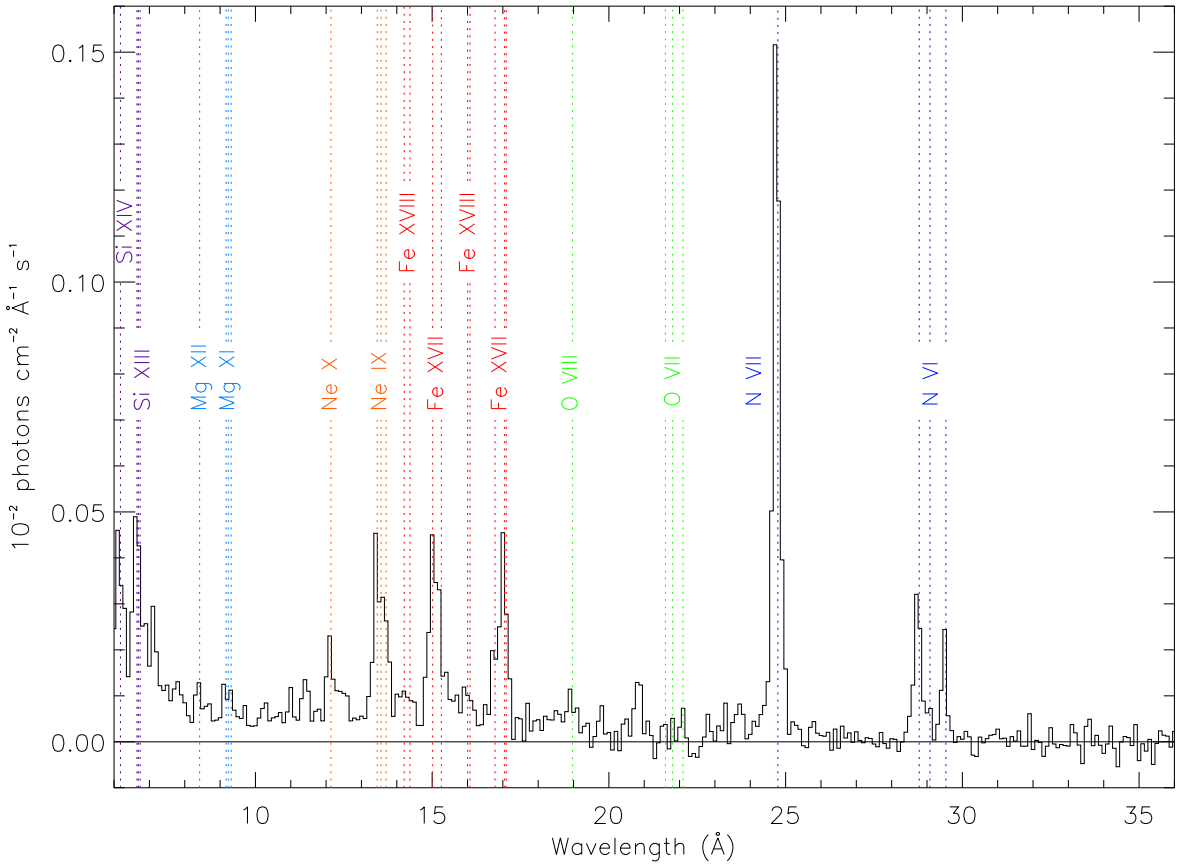, width=4.5in,angle=90}
  \end{center}
\caption{First order RGS spectrum of \protect\ec. It has been background
subtracted and corrected for effective area.}
\label{rgsspectrum}
\end{figure*}

The RGS spectrum is shown in Figure~\ref{rgsspectrum}. It is background
subtracted and corrected for effective area, as described below. It shows
emission lines from helium-like and hydrogen-like neon and nitrogen, and also
\ion{Fe}{17} and\ion{}{18}. These lines all originate from the nebula, in
contrast with the higher energy emission from the star. There is no obvious
emission from oxygen, from which one would expect prominent emission lines,
given the range of temperatures implied by the presence of the other emission
lines. There is also no obvious emission from charge states of iron higher than
\ion{Fe}{18}. \ion{Fe}{19} would be harder to see, as the brightest
emission line would lie at $\sim13.5$~\AA, which would be
indistinguishable from \ion{Ne}{9} at this resolution; however, if a
substantial amount of \ion{Fe}{20} emission was present, a strong emission
line would be present at $\sim12.8$~\AA. There is also no evidence
for detectable continuum emission. Thermal bremsstrahlung is less prominent
relative to line emission at temperatures around 0.5 keV than at higher
temperatures. Thus, from inspection it is clear that although the plasma is
not isothermal, the range of temperatures present is limited, and also that
the abundance of nitrogen is supersolar while that of oxygen is subsolar.

There are also emission lines from helium-like and hydrogen-like magnesium and
silicon, but these originate from the point source, with the possible
exception of \ion{Mg}{11}. This is known from the \ch\ HETGS spectrum
\citep{cor01}, which does not include emission from the nebula and shows
emission from all of these lines. The RGS cross-dispersion profiles of these
emission lines are consistent with point-like emission. The cross-dispersion
profile of \ion{Mg}{11} is also consistent with extended emission, so its
origin is unclear. Those emission lines are physically associated with
emission extending to much higher energies than are accessible with RGS, so we
do no attempt to model them in the analysis presented below.

Two main complications are encountered in the analysis of these RGS
data. First, the X-rays come from an extended source, and second, there
is a substantial background flux of soft X-rays which are diffuse or
unresolved, and which originates over essentially the entire spatial field of
view of RGS. These diffuse/unresolved X-rays presumably come from the many
OB stars in the field of view, and from truly diffuse, hot, shocked gas. The
diffuse emission affects all wavelengths, but the most severe confusion occurs
in the wavelength range $\sim10-20$ \AA. This is because the diffuse gas and
unresolved point sources have substantial iron L-shell and \ion{O}{8} emission,
and also because the column density is high enough to absorb most of the
diffuse emission at long wavelengths.

\begin{figure}[ht]
  \begin{center}
    \psfig{file=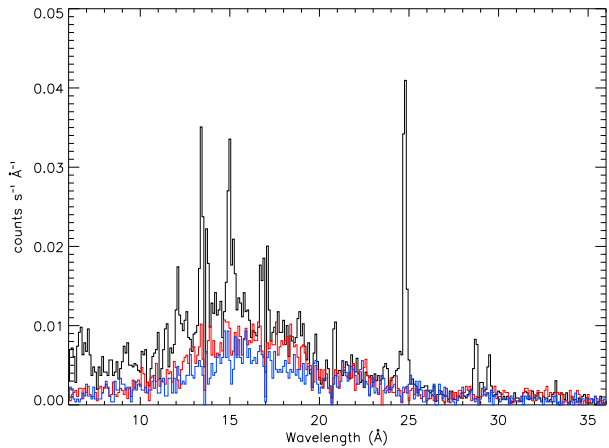, width=2.3in, angle=90}
  \end{center}
\caption{Spectra of the source (black) and background (blue and red) regions
in RGS1. The largest discrepancies between the two background spectra occur
around $18-19$ \AA.}
\label{spec_bkg}
\end{figure}

Because the soft X-ray emission is extended, the spectral and spatial
information are inherently coupled. A Monte Carlo simulation can be a useful
data analysis technique in this case \citep{pet02}.
This is not a feasible approach for this observation because there is no way
to reproduce the diffuse/unresolved background component with a simple
model. Background subtraction is also an issue, because the background point
sources and diffuse emission could have different spectra, and are not
uniformly distributed on the sky. In practice, the background spectrum
produced by a segment of the data offset from \ec\ in cross-dispersion
coordinates is fairly constant. The RGS~1 count rate spectra from the source
and background cross-dispersion regions are shown in figure~\ref{spec_bkg} to
demonstrate this, although there are differences in the crucial region around
\ion{O}{8} Ly $\alpha$. The adopted procedure is to produce a background
subtracted spectrum which includes an assessment of the systematic error
caused by the background subtraction. We estimate the potential systematic
error to be 25\% of the background strength in a given wavelength bin. This
systematic error is most important in low flux lines with high background,
e.g. \ion{O}{7} and\ion{}{8}.

Using this method, we then measure emission line strengths by taking
the total flux in the neighborhood of the line, taken to be within 0.3 \AA\ of
the rest wavelength. Because the bremsstrahlung
continuum emission is weak compared to line emission in the inferred
temperature range, this does not produce a substantial overestimate of the
line fluxes. For a few important emission lines, one expects strong emission
from other ions at about the same wavelength, which would lead to an
overestimate of the line strength. The most important correction is to
\ion{Ne}{10} Lyman $\alpha$, which is at roughly the same wavelength as the
strongest 4d-2p transitions in \ion{Fe}{17}. A correction is made based on the
observed 3-2 transitions in \ion{Fe}{17}. 

A Monte Carlo simulation of the RGS effective area is used together with the
measured source counts to obtain the emission line fluxes. Archival \ch\
ACIS-I imaging observations of \ec\ are used to provide a spatial
distribution for soft photons. The sky coordinates of photons with energy
below 1.2 keV are used as a spatial event list for the Monte Carlo
simulation. We assume that the spatial distribution does not vary as a
function of energy, and that the exposure map is approximately constant over
the $\sim$ 1 arcmin size of the nebula. Actual exposure variations are at or
below the 1\% level.

\begin{figure}[ht]
  \begin{center}
    \psfig{file=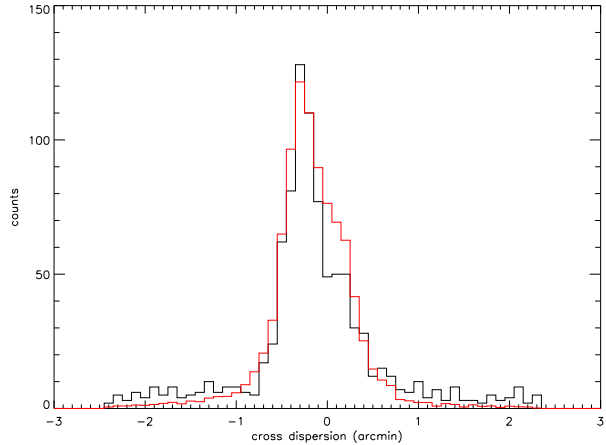, width=2.3in, angle=90}
  \end{center}
\caption{Cross-dispersion profile of \ion{N}{7} Lyman $\alpha$. The black line
is the data, and the red line is the Monte Carlo using the \protect\ch\
image.}
\label{n7xdsp}
\end{figure}

\begin{figure}[ht]
  \begin{center}
    \psfig{file=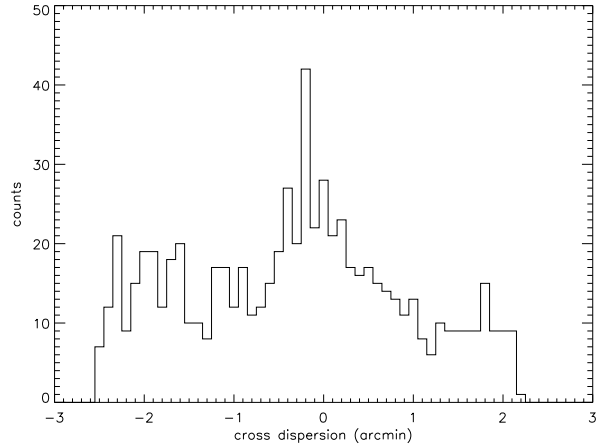, width=2.3in, angle=90}
  \end{center}
\caption{Cross-dispersion profile of \ion{O}{8} Lyman $\alpha$. }
\label{o8xdsp}
\end{figure}

\begin{figure}[ht]
  \begin{center}
    \psfig{file=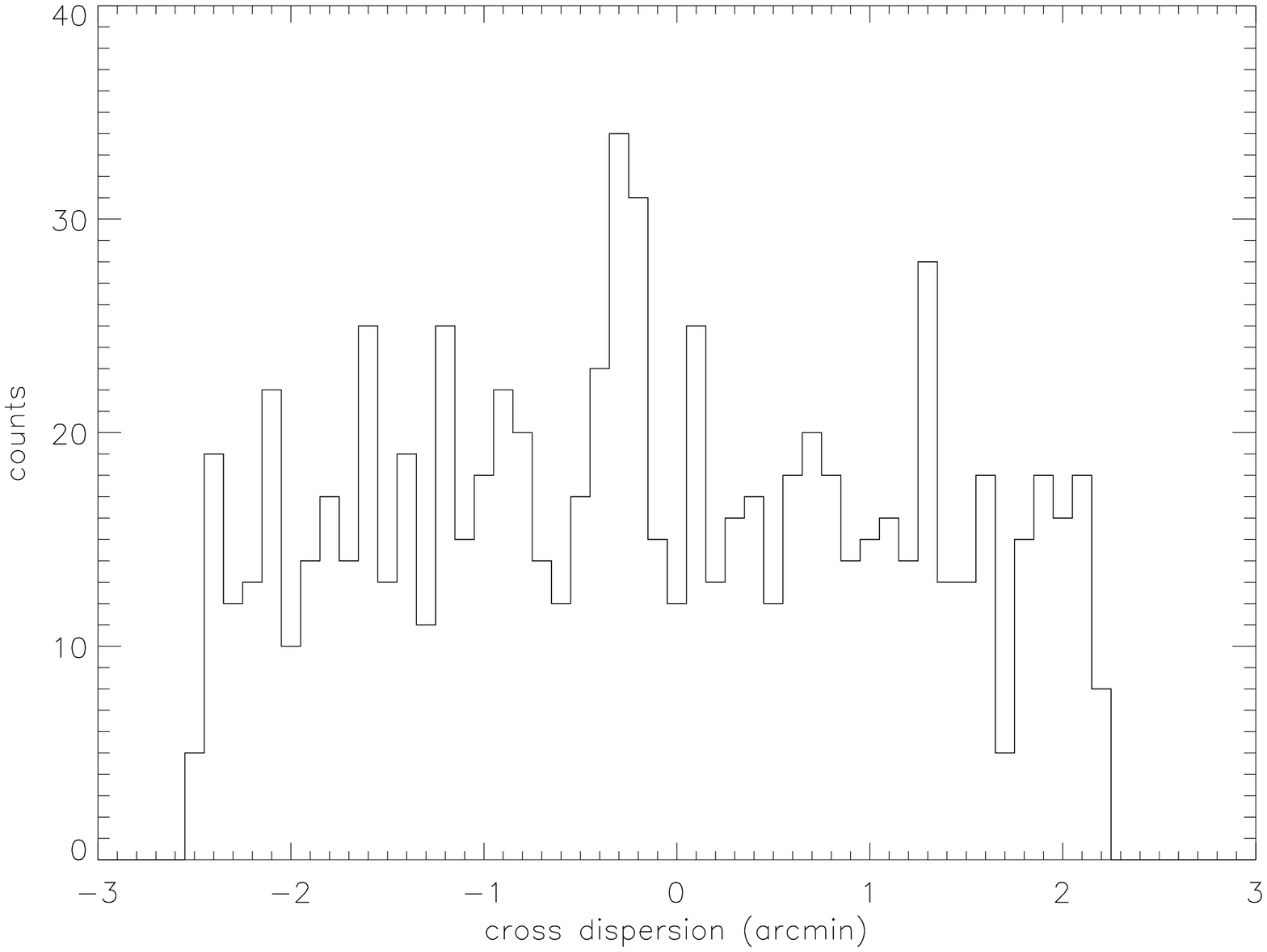, width=2.3in, angle=90}
  \end{center}
\caption{Cross-dispersion profile of \ion{O}{7} He $\alpha$. }
\label{o7xdsp}
\end{figure}

Figure~\ref{n7xdsp} shows the RGS 2 cross-dispersion image of the \ion{N}{7}
Lyman $\alpha$ line, including all photons within 0.3 \AA\ of the rest
wavelength, plotted together with the Monte Carlo cross-dispersion image using
the \ch\ data. The profiles are slightly different, which indicates that there
is some variation in the image of the nebula at different
temperatures. For comparison, Figures~\ref{o8xdsp} and \ref{o7xdsp} show the
cross-dispersion images of \ion{O}{8} Ly $\alpha$ and \ion{O}{7} He $\alpha$
respectively. A more detailed discussion of the significance of these
features follows below.


Table~\ref{TabFlux} gives the measured line fluxes for the major emission line
complexes in the RGS spectrum of \ec. It also gives the intrinsic line fluxes,
which correct for interstellar absorption, and in the case of \ion{Ne}{10}
Lyman $\alpha$ for the \ion{Fe}{17} 4d-2p lines.

Because the continuum flux in the soft X-ray spectrum is negligible, it is not
possible to measure neutral edge strengths or the equivalent neutral hydrogen
column density with RGS. The column density used in the X-ray literature
for \ec\ is $N_{\mathrm{H}}=2\times 10^{21} \, \mathrm{cm^{-3}}$
\citep{sew79}. This value comes from three sources. The first is a fit to the
low resolution IPC spectrum, giving a value of $2\times 10^{21} \,
\mathrm{cm^{-3}}$. The second is the \cite{sav77} measurement of UV \ion{H}{1}
and $\mathrm {H_{2}}$ absorption to a somewhat nearby star, HD 92740 (= WR
22), for which N(\ion{H}{1} + $\mathrm {H_{2}}$)$=1.8\times 10^{21} \,
\mathrm{cm^{-3}}$. The last is the conversion of the optical extinction to
neutral hydrogen column density using the relations of \cite{gor75} and
\cite{ryt75}. Stars near to \ec\ have $E_{\mathrm{B-V}}=0^{\mathrm{m}}.4$
\citep{fein73}, so the column density obtained is $2.7\times 10^{21}
\,\mathrm{cm^{-3}}$. Other measurements of $E_{\mathrm{B-V}}$ to nearby
stars yield similar results to \cite{fein73} \citep{her76, for78}. 

Of these methods, we cannot rely on the first, as the fitting procedure is
degenerate even for the high resolution RGS spectrum, and the second is also
unsatisfactory, as HD 92740 is too far away from \ec\ to expect the column
density to be the same. The third method is also problematic, as the
extinction to Tr 16 is anomalous and variable. The extent to which it is
anomalous is controversial, with different works obtaining values of $R =
A_{\mathrm{V}}/E_{\mathrm{B-V}}$ ranging from 3.2 to 5.0 for different stars
in Tr 16 and in the vicinity of \ec\ \citep{fein73, her76, for78, tur80,
the83, tap88, the95}. This implies that the relations of
\cite{gor75} and \cite{ryt75} underestimate the total column density, since
they take $R \sim 3.1$. However, these relations should at least provide a
lower limit to $N_{\mathrm{H}}$. In their review, \cite{dav97} adopt a value of
$A_{\mathrm{V}} = 1.7$ for the interstellar (non-circumstellar) extinction to
\ec. Using the \cite{gor75} relation, we obtain $N_{\mathrm{H}}=3.7\times
10^{21} \, \mathrm{cm^{-3}}$. Also, as an alternative to the \cite{sav77}
measurement, we can use the \cite{dip94} measurement of the \ion{H}{1} column
towards HD 303308, which is much nearer to \ec\ than HD 92740. They find
$N_{\mathrm{H}}=2.8\times 10^{21} \,\mathrm{cm^{-3}}$. This can also be
considered a lower limit, as a substantial fraction of the hydrogen may be
ionized. For simplicity we will take the equivalent hydrogen column density to
be $N_{\mathrm{H}}=3.0\times 10^{21} \, \mathrm{cm^{-3}}$, and assess the
systematic effects of a higher column on the temperature distribution and
abundance measurements.

There is no evidence for emission from \ion{O}{7} He~$\alpha$. Although there
is no strong emission line corresponding to \ion{O}{8} Lyman~$\alpha$, there
is a marginal detection of flux at this wavelength, and a small line-like
feature. Because \ion{O}{8} Ly~$\alpha$ is so weak in this spectrum, and
because it is one of the stronger features in spectra of typical O-type stars
and collisionally ionized plasmas in general, the question of
contamination by nearby sources is important to address. It is possible that
some or all of the observed \ion{O}{8} Ly~$\alpha$ flux is attributable to a
nearby star. HD 303308 is the brightest object close enough to cause
confusion. If this star was the source of the apparent \ion{O}{8} feature, it
would have an apparent wavelength of 18.9 \AA\ (slightly blueshifted),
and it would be about 1 arcmin east of the bright knot seen in the southwest
of the \ch\ image. The observed RGS feature is inconsistent with this
requirement. In addition, the \ion{O}{8} flux obtained from a two temperature
thermal plasma model fit to the EPIC-MOS spectrum is $10^{-5} \,
\mathrm{photons \, cm^{-2} \, s^{-1}}$, compared with $4 \pm 2 \times 10^{-5}$
for \ec. Thus, it seems unlikely that any point source can account for the
\ion{O}{8} feature in \ec. However, \ion{O}{8} Ly $\alpha$ is only detected at
the 2$\sigma$ level, so the detection is not very secure.

We do not use the data to obtain an upper limit to the \ion{C}{6}
emission. The high column density would make any upper limit a weak one, and
the fact that we expect a very low carbon abundance means that the more easily
measurable oxygen abundance provides a stronger physical constraint on the
nucleosynthetic signatures of the CNO cycle.

\begin{figure*}[ht]
  \begin{center}
    \psfig{file=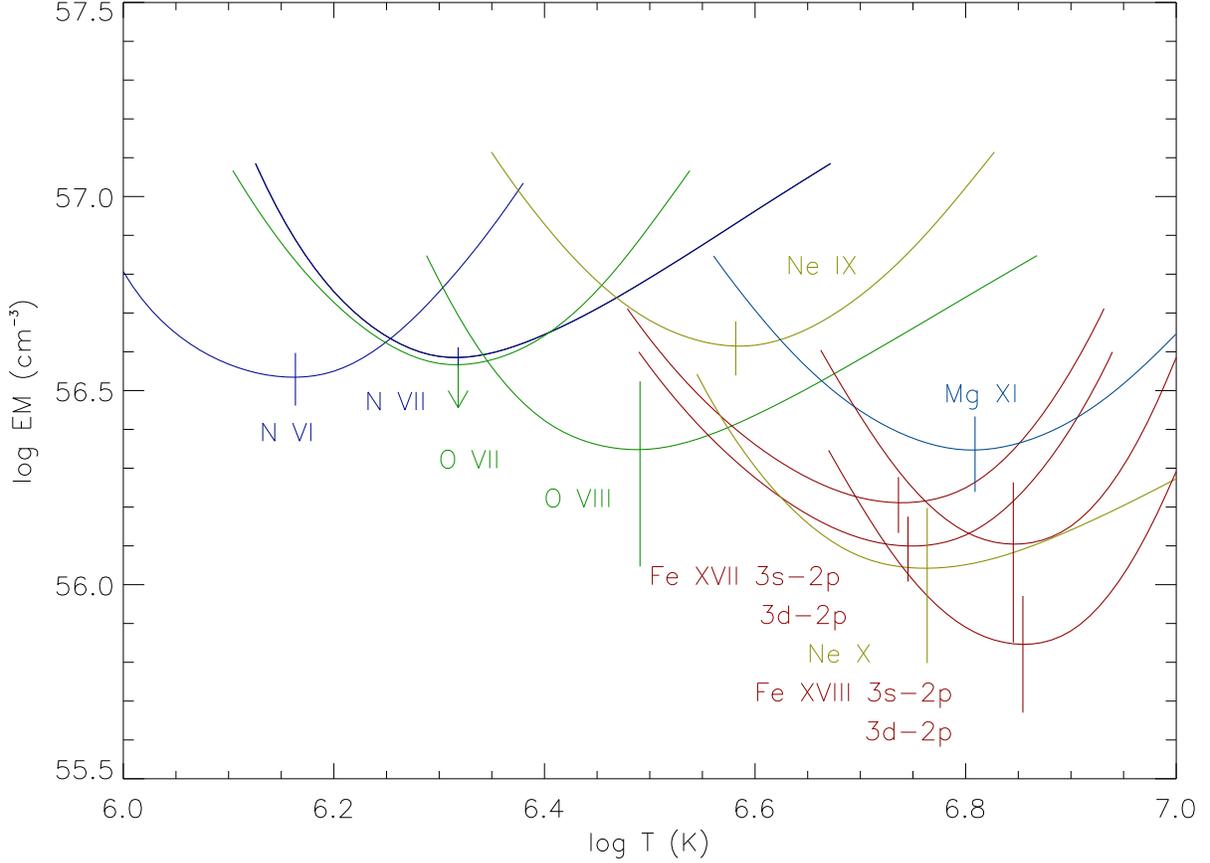, width=4.5in, angle=90}
  \end{center}
\caption{Inferred emission measure distribution. The abundance of nitrogen is
set to the sum of solar C+N+O, while oxygen is set so that \ion{O}{7} is
consistent with \ion{N}{7}. Other elements have solar abundances.}
\label{etadem}
\end{figure*}

In Figure~\ref{etadem} we plot the emission measure inferred for individual
emission line complexes assuming a single temperature. For each ion, the error
bar is marked at the temperature of maximum line strength. The abundances are
taken to be solar \citep{ag89}, except nitrogen and oxygen. The abundance of
nitrogen is assumed to be 11 times solar, which is approximately the sum of
the solar abundances of carbon, nitrogen, and oxygen. The abundance of oxygen
is chosen to have a value such that \ion{O}{7} and \ion{N}{7} are consistent,
as discussed below. It is clear that a single temperature
cannot account for the different species observed. Furthermore, the lack of
emission from \ion{Fe}{20} shows that there cannot be a substantial amount of
plasma at temperatures above 0.6 keV (7 MK). The presence of \ion{N}{6}
emission shows that there must be emission from temperatures down to at least
about 0.15~keV (1.7 MK), but emission at lower temperatures is unconstrained;
there are no spectral features we would expect to see if lower temperatures
were present, given the probable low carbon abundance and high column density.
 

Because the strength of the \ion{N}{7} and \ion{O}{7} lines have very similar
temperature dependence, especially near the temperature of maximum line
strength ($T_{\mathrm{m}}$), one may derive an upper limit to the abundance
ratio by setting the oxygen abundance such that the curves in the plot are
consistent near $T_{\mathrm{m}}$. This is possible because we know that there
is no emission from high temperatures, where the relative flux of \ion{N}{7}
and \ion{O}{7} has a substantial temperature dependence. The same is true for
\ion{Fe}{17} and \ion{Ne}{10}. We measure a lower limit $\mathrm{N/O} > 9$,
while Fe/Ne is consistent with the solar ratio \citep{ag89} to 0.1 dex. The
error introduced by directly comparing ion emission measures assuming a single
temperature is less than 0.1 dex. If we allow that the column density assumed
could be $4 \times 10^{21}\, \mathrm{cm^{-2}}$, the lower limit to N/O
becomes $\mathrm{N/O} > 8$. Although it is possible that the column density is
higher than that, this uncertainty clearly cannot affect the N/O lower limit
very strongly. 

Alternatively, if we take the \ion{O}{8} detection at face value and fix the
oxygen abundance by requiring that the differential emission measure should
not have a dip at $\mathrm{log}(T)=6.5$, the N/O ratio is increased by about
0.2 dex from our lower limit, to $\mathrm{N/O} = 14$.

The appearance of the ion emission measure plot is moderately affected by the
set of reference abundances chosen. The most recent work on solar abundances
\citep{gs98,h01,ap01,ap02} have substantial variations relative to each other
and \cite{ag89} in the abundances of CNO and Fe. These variations are of order
0.1 dex. Given the levels of uncertainty in the measurements themselves and
in the abundances, the plot is consistent with a differential emission measure
which is flat from $\mathrm{log}(T)=6.0-6.6$, declining substantially to
$\mathrm{log}(T)=6.8$. In any case, the measurement of the ratio N/O is
unaffected by the choice of reference abundances.

The shape of the emission measure distribution would be affected if the column
density were substantially higher than $4 \pm 1 \times 10^{21}\,
\mathrm{cm^{-2}}$. Rather than looking flat, with a dropoff at high
temperatures, it would be decreasing with increasing temperature.

\ion{Mg}{11} has also been included in the plot. For this point to be
consistent with the rest of the plot, we would have to take the abundance of
magnesium to be at least 0.3 dex higher than iron and neon. While this is not
out of the question, there is no good reason to have an overabundance of only
magnesium (which would have no relation to CNO cycle abundance changes if it
were real). The most likely explanation is that \ion{Mg}{11} emission comes
from the star rather from the nebula. There is also evidence for a feature at
about 7.85 \AA, the wavelength of \ion{Mg}{11} He $\beta$. This
feature is strong compared to \ion{Mg}{11} He $\alpha$, as would be expected
for the high column density observed in the spectrum of the star. This is also
similar to the \ion{Mg}{12} Ly $\alpha$ to $\beta$ ratio observed with {\it
Chandra} \citep{cor01}.


\begin{table*}[ht]
\begin{center}
\caption{Measured fluxes for prominent emission line complexes with rest
wavelengths in {\AA}. The intrinsic flux of \ion{Ne}{10} Ly $\alpha$ has been
corrected for blending with 4-2 transitions of \ion{Fe}{17}.}
\begin{tabular}{ccc}
\hline
\hline
Line & Flux & Intrinsic Flux \\
  & ($10^{-4} \, \mathrm{photons\, cm^{-2}\, s^{-1}}$) & ($10^{-4} \, \mathrm{photons\, cm^{-2}\, s^{-1}}$) \\
\hline
\ion{Mg}{11} He $\alpha$   & $0.59\pm0.13$ & $0.84\pm0.18$ \\
\ion{Ne}{10} Ly $\alpha$    & $0.71\pm0.14$ & $0.73\pm0.31$ \\
\ion{Ne}{9} He $\alpha$   & $1.76\pm0.28$ & $4.47\pm0.71$ \\
\ion{Fe}{18} 3d-2p      & $0.51\pm0.17$ & $1.23\pm0.41$ \\
\ion{Fe}{18} 3s-2p      & $0.43\pm0.19$ & $1.47\pm0.65$ \\
\ion{Fe}{17} 3d-2p       & $1.84\pm0.35$ & $5.32\pm1.01$ \\
\ion{Fe}{17} 3s-2p       & $1.95\pm0.32$ & $8.15\pm1.34$ \\
\ion{O}{8} Ly $\alpha$  & $0.4\pm0.2$   & $2.2\pm1.1$ \\
\ion{O}{7} He $\alpha$   & $<0.4$        & $<5.6$ \\
\ion{N}{7} Ly $\alpha$   & $4.02\pm0.25$ & $36.7\pm2.3$ \\
\ion{N}{6} He $\alpha$    & $1.55\pm0.24$ & $52.4\pm8.1$ \\

\hline
\label{TabFlux}
\end{tabular}

\end{center}
\end{table*}


\section{Discussion}
\label{dis}

There are two main results from the analysis of the RGS spectrum of
\ec. The first is a constraint on the range of temperatures in the nebula
($0.15-0.6\,\mathrm{keV}$), which allows us to infer shock velocities for the
expansion of the ejecta into the surrounding medium. The second is a lower
limit on the nitrogen to oxygen abundance ratio ($\mathrm{N/O} > 9$). This
allows us to constrain the evolution of \ec.

\subsection{Temperature distribution}

The upper end of the temperature distribution is strongly constrained by the
Fe L-shell spectrum. The lack of measureable emission from charge states
higher than \ion{Fe}{18} rules out the presence of appreciable quantities of
gas above $\sim\,0.6\,\mathrm{keV}$. The lower end of the distribution
appears to be flat, based on the emission from \ion{Ne}{9}, \ion{N}{7} and
\ion{N}{6}. However, there are no other potentially observable spectral lines
originating from ions that exist at lower temperatures than \ion{N}{6}, so
the emission measure distribution cannot be constrained below about 0.2
keV. \cite*{dav82,dav86} find UV emission lines from \ion{N}{1} through
\ion{N}{5} in the spectra of the ejecta, so there is certainly a range of
temperatures present. As noted in the previous section, a substantial
difference between the assumed absorption and the real absorption could change
the overall shape of the emission measure distribution, especially at low
temperatures.

\cite*{wei01a} attempt to correlate the observed projected velocity of optical
blobs which are spatially coincident with X-ray emission using Hubble Space
Telescope and ROSAT HRI images. The velocities they find in the brightest
X-ray regions would produce plasma at temperatures an order of magnitude
higher than observed, assuming that the ejecta was colliding with a stationary
ISM. Of course, \ec\ should be surrounded by a wind blown bubble out to much
larger radii than 0.3~pc (the radius of the X-ray nebula), and the material
inside the bubble should be streaming outward. It seems likely that the
observed shock temperature reflects the velocity at which the X-ray emitting
ejecta are overtaking the previously emitted stellar wind. The temperature
range $0.15-0.6\,\mathrm{keV}$ implies a shock velocity range of
$300-700\,\mathrm{km\,s^{-1}}$. If the X-ray emitting ejecta date from the
great eruption of 1843, then the rough expansion velocity for a free expansion
is $\sim\,\mathrm{0.3\,pc\,/\,150\,yr\,=\,2000\,km\,s^{-1}}$, so the velocity
of the stellar wind before the great eruption was
$\sim\,\mathrm{1500\,km\,s^{-1}}$.

\subsection{Abundance measurements}

The N/O ratio observed in the ejecta has implications for the evolution
of \ec. It is clearly a signature of CNO processing, and the degree of
conversion of oxygen to nitrogen observed in the ejecta is high.

All massive main-sequence stars burn hydrogen on the CNO cycle. Its
nucleosynthetic signatures are the conversion of most of the catalytic carbon
and oxygen to nitrogen, and the burning of H into He. For CNO processed
material to be observed on the surface of these stars, or in their ejecta, it
must be transported there from the core. 

Previous measurements of N/O in \ec\ give similar but generally less
constraining results than our RGS measurements. Optical and UV spectroscopy of
the S condensation (corresponding spatially roughly to the brightest X-ray
knot \citep{sew01}) shows that most CNO is nitrogen and that the helium mass
fraction is $0.40\pm0.03$ \citep{dav82,dav86}. A quantitative measurement of
the CNO abundance ratios is not made because of their dependence on ionization
and thermal structure, and also because some oxygen and carbon may be in solid
grains. It should be noted that the measured value of the helium mass fraction
may be systematically too low if the ionization balance of helium was not
properly modelled.

More recent measurements of the abundances in the S condensation have been
made by \cite{duf97} with HST-FOS. They report CNO and He abundances for the
S2 and S3 ``sub-condensations'', respectively, of $\mathrm{[N/O]}>1.72,1.75$,
$\mathrm{[N/C]}>1.95,1.85$, and $Y = 0.39,0.42$. They did detect weak oxygen
and carbon lines, but treated them as upper limits due to potential
contamination from the foreground \ion{H}{2} region. However, they also find
that preliminary analysis of the S1 and S4 sub-condensation spectra show much
lower N and He enrichment, with correspondingly lower N/O and N/C ratios. 

Previous X-ray observations \citep{tsu97,cor98,sew01,wei01b} have shown the
presence of a strong \ion{N}{7} Ly~$\alpha$ feature in the spectrum, but the
CCD spectra lacked the resolution to strongly constrain the \ion{O}{7} and
\ion{O}{8} features. Our measurement of N/O is not limited by the spectral
resolution of RGS, but rather by source/background contamination, and the
observed line strength is not influenced by the formation of dust grains or
large uncertainties in the temperature distribution of the plasma.

Recent HST-STIS long-slit spectroscopy of the central star have obtained a
lower limit of $\mathrm{N/O}\gax 1$ \citep{hil01}. This is a conservative
interpretation of the data; the lower limit could easily be taken to be an
order of magnitude higher. On the other hand, UV spectra taken with HST-GHRS
show evidence for moderate carbon depletion which may be inconsistent with the
level of depletion found in the ejecta \citep{lam98}. In light of recent
work indicating that \ec\ may be a binary system \citep{dam96,dam00}, the
apparent contradiction in the stellar and nebular abundances is taken to be an
indication that the star producing the carbon features is actually the
secondary (assuming the star that produced the nebula is the
primary). \cite{wal99} points out that there are several difficulties with
this conclusion, the most obvious being the concealment of the luminous blue
variable (LBV) primary.

Spectroscopic measurements of the abundances of the central star cannot
invalidate the RGS abundance measurements, but the binary scenario requires us
to treat the nebular abundances with some care. It is unlikely that both
members of a binary system could contribute substantially to the ejecta around
\ec, but it is possible, in principle, that the ejecta from the primary could
mix with the wind of the secondary. If both stars had a high N/O ratio, but
substantially different helium abundances, then it would be possible
to misinterpret the significance of the nebular abundances. However, this is
not a likely scenario, so we make the simplest assumption, which is that the
observed nebular abundances reflect the current surface abundances of the
primary. 

The signatures of CNO processing have been observed in various types of hot
stars, including OBN stars, blue supergiants, and LBVs \citep{mae95}. However,
CNO processed material is not observed on the surface of all hot stars, and
the amount of processed material observed spans a wide range. The fact that
N/O is so high in the ejecta of \ec\ is strongly constraining, regardless of
the mechanism responsible for mixing.

The two most plausible mechanisms which could have resulted in the measured
abundances in the ejecta of \ec\ are :
1.) \ec\ is on the main sequence and is rotating. This rotation has caused
very thorough mixing.
2.) \ec\ is in a post-red-supergiant blue supergiant phase, and the CNO
abundance ratios are a result of the onset of convection in the envelope
during the red supergiant phase.
We refer in particular to the discussion in \cite{lam01}, which deals with the
same question in the case of other LBV nebulae. We can use the measured N/O
ratio in conjunction with the He abundance of \cite{dav86} to assess the
plausibility of these two mechanisms.

Using Figure 3 of \cite{lam01} for the case of an $85\, \mathrm{M_{\odot}}$
star with $\mathrm{Z}\,=\,0.02$, we find that for $\mathrm{log(N/O)}>1.0$,
$\mathrm{log(He/H)}>-0.3$, or $\mathrm{Y}>0.67$. This simply reflects the
fact that although a high surface ratio of N/O can be obtained in the red
supergiant phase, this can only happen if the star lost enough of its envelope
on the main sequence to allow core processed material to dominate the
resulting composition. This value of Y is not consistent with the \cite{dav86}
measurement of $\mathrm{Y}\,=\,0.4$, although a conservative assessment of the
possible errors, particularly in measuring the helium mass fraction, does not
allow us to rule out that \ec\ could be a post-red-supergiant object. 

\cite{mey00} make predictions for the abundances of rotating massive
stars. Their $\mathrm{Z}\,=\,0.02$ model with an initial rotation velocity of
$300\,\mathrm{km\,s^{-1}}$ and a mass of $120\,\mathrm{M_{\odot}}$ predicts
$\mathrm{Y_{s}}=0.89$ and $\mathrm{N/O}=45.4$ at the end of H-burning. While
this value of Y is also not consistent with the observed value, in this case Y
will clearly be lower earlier in the life of the star, whereas if the mixing is
efficient enough, N/O will already be high enough to be consistent with the
measured lower limit. This is an important point; the conversion of oxygen to
nitrogen in CNO burning is considerably slower than the conversion of carbon
to nitrogen. If rotational mixing is responsible for the
observed abundances, the mixing timescale must be short compared to the
evolutionary timescale. As pointed out in \cite{mae87}, the ratio of the
mixing timescale to the main sequence lifetime in rapidly rotating stars is
indeed expected to decrease with increasing mass. 

\subsection{Summary}

We have analyzed XMM-Newton X-ray spectra of \ec. The EPIC spectral data from
the star are consistent with past observations by {\it ASCA} and \ch. The data
are not consistent with an isothermal plasma, but require at least two
temperatures. 

The RGS spectra show that the nebula is nonisothermal and
has strongly non-solar CNO abundances. The temperature range in the
nebula is $0.15\,-\,0.6\,\mathrm{keV}$. If this is interpreted as a shock
velocity, it corresponds to $300\,-\,700\,\mathrm{km\,s^{-1}}$. We find a
lower limit of $\mathrm{N/O}\,>\,9$, which is indicative of very thorough
mixing in the envelope of \ec. Taken with previous measurements of the surface
helium abundance $\mathrm{Y}\,=\,0.4$, this implies that \ec\ is a
main-sequence object with some strong mixing mechanism at work, although it
does not decisively rule out the possibility that it is a post-red-supergiant object.

\begin{acknowledgements}

We acknowledge useful comments from the anonymous referee, which substantially
improved the quality of this paper. MAL thanks J. R. Peterson for useful
conversations and for the development of features relevant to this analysis
for the RGS Monte Carlo. The Columbia University group is funded by NASA.

\end{acknowledgements}

\end{document}